\documentclass[onecolumn,showpacs,amsmath,amssymb,prd,nofootinbib]
{revtex4}
\def\sss{\scriptscriptstyle}
\def\^#1{^{\sss #1}}
\def\_#1{_{\sss #1}}
\def\beq{\begin{equation}}
\def\eeqno#1{\label{#1}\end{equation}}
\def\ten#1#2{^{\sss#1}_{\sss#2}}

\def\kms{{\rm km~s^{-1}}}

\def\gcmt{{\rm gr~cm^{-3}}}

\def\msun{M\_{\odot}}
\def\az{a\_{0}}

\def\l0{\ell\_{0}}

\def\rar{\rightarrow}

\def\l{\lambda}

\def\rp{\rho_p}

\def\f{\phi}
\def\fN{\phi\^N}
\def\fNh{{\hat\phi}\^N}

\def\gfb{\grad\fb}

\def\z{\zeta}
\def\e{\eta}

\def\r{\rho}
\def\rh{\hat\rho}
\def\m{\mu}
\def\mt{\tilde\mu}
\def\n{\nu}
\def\Up{\Upsilon}
\def\C{\Gamma}

\def\L{\mathcal{L}}

\def\M{\mathcal{M}}

\def\Th{\hat\mathcal{T}}

\def\D{\Delta}

\def\d{\delta}

\def\a{\alpha}
\def\b{\beta}
\def\c{\gamma}
\def\d{\delta}
\def\eps{\epsilon}
\def\vr{{\bf r}}

\def\vR{{\bf R}}

\def\vF{{\bf F}}

\def\vv{{\bf v}}

\def\vu{{\bf u}}

\def\vk{{\bf k}}

\def\va{{\bf a}}

\def\vF{{\bf F}}

\def\grad{\vec\nabla}
\def\div{\vec \nabla\cdot}
\def\gf{\grad\phi}
\def\fpg{4\pi G}

\def\gmn{g\_{\m\n}}
\def\Gmn{g\^{\mu \nu}}

\def\Gab{g\^{\alpha\beta}}

\def\hGmn{\hat g\^{\m\n}}
\def\hgmn{\hat g\_{\m\n}}
\def\hgh{\hat g\^{1/2}}
\def\gh{g^{1/2}}

\def\gft{\grad\tilde\f}
\def\fh{\hat\f}
\def\gfh{\grad\fh}

\def\ft{\tilde\f}
\def\fb{\bar\f}
\def\gfb{\grad\fb}
\def\hC{\hat\C}
\def\cd#1{{}_{\sss;#1}}
\def\mcd#1{{}_{\sss:#1}}

\def\T{\mathcal{T}}
\def\Tmn{\T\_{\m\n}}
\def\Th{\hat{\mathcal{T}}}
\def\hTmn{\Th\_{\m\n}}
\def\emn{\eta\_{\m\n}}
\def\rph{\hat\rp}

\def\rz{\r\_b}
\def\vvz{\vv\_b}
\def\pz{p\_b}
\def\T{\mathcal{T}}
\def\Tmn{\T\_{\m\n}}
\def\Th{\hat{\mathcal{T}}}
\def\hTmn{\Th\_{\m\n}}

\def\oot{\frac{1}{2}}

\begin{document}

\title{Cosmological fluctuation growth in bimetric MOND}
\author{Mordehai Milgrom }
\affiliation{Department of Particle Physics and Astrophysics,
Weizmann Institute, Rehovot 76100, Israel}
\begin{abstract}
I begin to look at the growth of weak density inhomogeneities of
nonrelativistic matter, in bimetric-MOND (BIMOND) cosmology. Far
from making an exhaustive study, I concentrate on one attractive
cosmological scenario, which employs matter-twin-matter-symmetric
versions of BIMOND, and, furthermore, assumes that, on average, the
universe is symmetrically populated in the two sectors. MOND effects
are totally absent in an exactly symmetric universe, apart from the
significant possible appearance of a cosmological constant,
$\Lambda\sim(\az/c)\^2$. MOND effects--local and cosmological--do
enter when density inhomogeneities that differ in the two sectors
appear and develop. MOND later takes its standard form in systems
that are islands dominated by pure matter, as are presumably the
well scrutinized systems such as galaxies. I derive the
nonrelativistic (weak-field-slow-motion) equations governing
small-scale fluctuation growth. The equations split into two
uncoupled systems, one for the sum, the other for the difference, of
the fluctuations in the two sectors. The former is governed strictly
by Newtonian dynamics, and describes standard growth of
fluctuations. The latter is governed by MOND dynamics, which entails
stronger gravity, and nonlinearity even for the smallest of
perturbations. These cause the difference to grow faster than the
sum, leading to anticorrelated perturbations, conducing to
matter-twin-matter segregation (which continues for high
overdensities). The nonlinearity also causes interaction between
nested perturbations on different scales. Because matter and twin
matter (TM) repel each other in the MOND regime, matter
inhomogeneities grow not only by their own self gravity, but also
through shepherding by flanking TM overdensities (and vice versa).
The relative importance of gravity and pressure in the MOND system
(analog of the Jeans criterion), depends also on the strength of the
perturbation. MOND gravity, which scales as the square root of the
difference perturbation, holds sway over pressure for any mass, for
weak enough perturbations. The development of structure in the
universe, in either sector, thus depends crucially on two initial
fluctuation spectra: that of matter alone and that of the matter-TM
difference. I also discuss the back reaction on cosmology of BIMOND
effects that appear as ``phantom matter'' (interpreted by some as
``dark matter''), resulting from inhomogeneity differences between
the two sectors.
\end{abstract}
\maketitle
\section{Introduction}
A full fledged treatment of cosmology and of structure formation in
the MOND paradigm is still wanting. Structure formation on an
expanding cosmological background, in nonrelativistic (NR) versions
of MOND, has been studied with $N$-body calculations, based on
various heuristic extensions of MOND theories designed to work for
isolated NR systems, e.g., in
\cite{sanders01,nusser02,stach01,knebe04,ll08}. Clearly, a definite
treatment of the problem has to be based on a fully relativistic
theory, encompassing both cosmology and structure formation. Much
work along such lines has been done in the framework of TeVeS--a
relativistic MOND theory propounded by Bekenstein \cite{bek04}
(Sanders's stratified framework \cite{sanders97} is a precursor, and
\cite{skordis09} is a recent review of TeVeS and its cosmology). For
example,
\cite{sanders05,dodelson06,skordisetal06,skordis06,skordis08,
zlosnik08} considered various aspects of cosmology at large, the
cosmic microwave background (CMB), and linear stages of structure
formation in TeVeS-like theories. Although very encouraging, these
results do not yet provide a fully satisfactory description of the
phenomena at hand.
\par
A new relativistic theory for MOND (called BIMOND, for ``bimetric
MOND'') was proposed recently based on a bimetric structure of
space-time \cite{milgrom09}. In this theory, gravity is described by
two metrics instead of the one  underlying general relativity (GR).
Ordinary matter, to whose sector belongs our direct ken, couples, as
in GR, only to one metric, call it the primary one. The other metric
serves as an auxiliary to which the primary metric couples. Such a
bimetric description of gravity has a long history, and, in
particular, has been discussed recently in connection with some
``dark substance'' phenomena (see \cite{dk02,bdg06,bdg07,banados09},
and references therein). The novelty in BIMOND is in the special
choice of interaction between the two metrics, which is particularly
germane in the context of MOND, and which reproduces MOND
phenomenology.
\par
In \cite{milgrom09}, I made preliminary comments on cosmological
solutions within BIMOND. Classes of cosmological solutions within
certain versions of BIMOND were also considered in \cite{cz10}. But,
there is still much left to be done on this problem.
\par
We are also still lacking any treatment of the important problem of
the growth of small perturbations. Ideally, one should treat the
problem in a fully relativistic framework, as done in
\cite{skordis06,skordis08,zlosnik08}, in the context of TeVeS-like
theories. Here I take the lighter task of considering only density
inhomogeneities of NR matter on scales smaller than
cosmological. This limited-scope formalism, which governs the growth
of perturbations under matter dominance, is still rather
illuminating, and shows a variety of phenomena potentially highly
pertinent to structure formation in our universe.
\par
I limit myself to a universes that on average is completely
symmetric in matter and twin matter (TM). Twin matter is the
putative matter that couples to the ``auxiliary'' metric as
(ordinary) matter couples to the ``primary'' one. At present, the
existence of TM is not required phenomenologically (in particular,
it is not, in fact cannot be, a sort of ``dark matter,'' as assumed
to explain galaxy dynamics); it is invoked mainly for aesthetic
reasons: With TM we can obtain a simple and symmetric world picture
that affords favorable cosmologies. In addition, as our present
analysis shows, symmetric presence of matter and TM has very
interesting consequences for structure growth. It leads to phenomena
that are not only greatly different from those in GR (without dark
matter), but also different from those in MOND without TM.
\par
Of course, the existence of TM would add crucial unknowns to our
cosmological world view. Its properties, composition,
self-interactions, amounts, etc., are not known (see
\cite{milgrom09} for more on this). To minimize these unknowns I
assume here that, with one exception, the TM sector is identical to
the matter sector in all respects, having the same properties
(masses, etc.), the same interactions within the sector
(electromagnetic, strong, etc.), the same interaction with its
metric (geodesic motion, etc.), and the same average amounts of its
different components present in the universe. As shown in
\cite{milgrom09} this symmetry leads to simple cosmological
solutions, identical with those of GR, albeit with a naturally
appearing cosmological constant (CC) of order $(\az/c)\^2$, where
$\az$ is the MOND constant.\footnote{So the successes of standard
Big Bang cosmology, such as nucleosynthesis, are retained. The
symmetry assumption entails, for example, having the same inflation,
and the same mechanism for baryogenesis in the two sectors.} The
only difference between the two sectors I allow for, is in the
amplitude and spectrum of the primordial inhomogeneities. These are
putatively ascribed to quantum fluctuations in the early Universe.
They could, thus, differ in the two sectors--at least in their exact
distribution, if not in their statistical properties--even with all
else being the same.
\par
Some preliminaries of the problem discussed here are alluded to in
\cite{milgrom10}, where I considered the problem of NR, small,
isolated matter-TM systems. Here, this problem is extended to the
evolution of perturbations on an otherwise homogeneous and matter-TM
symmetric (MTMS), expanding universe.
\par
In section \ref{bimond}, I briefly describe the aspects of BIMOND
and of its NR limit that we need in this paper. In section
\ref{problem}, I derive the equations controlling the evolution of
small departures from exact density equality in general NR systems.
This is applied to cosmological small inhomogeneities in section
\ref{cosmo}. In section \ref{back}, I describe succinctly how the
density differences back react on the cosmological equations,
introducing modifications that can act as cosmological ``dark
matter'' and ``dark energy'' (beside the CC). Section
\ref{discussion} is a discussion.
\section{\label{bimond}BIMOND recapped}
BIMOND is a class of relativistic theories of gravity that involve
two metrics, $\gmn$ and $\hgmn$. They are governed by an action of
the form
 \beq I=-\frac{1}{16\pi G}\int[\b\gh R
+ \a\hgh \hat R
 -2(g\hat g)^{1/4}\az\^2\M] d^4x
+I\_M(\gmn,\psi_i)+\hat I\_M(\hat g\_{\m\n},\hat\psi_i),
\eeqno{mushpa} where $R$ and $\hat R$ are the Ricci scalars of the
two metrics as appear standardly in the Einstein-Hilbert action of
GR, $G$ is Newton's constant, and I use units where $c=1$. The
interaction term $\M$ is a dimensionless, scalar function of the two
metrics and their first derivatives.
\par
Bimetric theories have been much discussed before; the novelty and
crux of BIMOND is in the choice of the interaction term. The key
observation, in the MOND context, is that with two metrics and their
first derivatives, we can construct tensors with the dimensions of
acceleration:
 \beq C\ten{\a}{\b\c}=\C\ten{\a}{\b\c}-\hC\ten{\a}{\b\c},
  \eeqno{veyo}
where $\C\ten{\a}{\b\c}$, $\hC\ten{\a}{\b\c}$ are the Levi-Civita
connections of the two metrics. Within the MOND paradigm, having the
acceleration constant, $\az$, at our disposal, we can construct from
these dimensionless scalars to serve as variables on which $\M$
depends. These scalars are obtained by contracting the dimensionless
tensors $\az\^{-1}C\ten{\a}{\b\c}$.\footnote{In the units where
$c=1$, $\az$ has dimensions of $length\^{-1}$. In standard units,
$\az$ is replaced everywhere by $\ell\^{-1}\equiv \az/c\^2$. Then in
the NR limit only $\az$ appears. In standard units,
$C\ten{\a}{\b\c}$ have dimensions of $length\^{-1}$; so the
dimensionless tensors to be used are $\ell
C\ten{\a}{\b\c}=(c\^2/\az)C\ten{\a}{\b\c}$.} It befits us to
concentrate on scalars that are quadratic in the $C\ten{\a}{\b\c}$,
which I do. In particular, the scalars constructed from  the
quadratic tensor
 \beq \Up_{\m\n}\equiv  C\ten{\c}{\m\l}C\ten{\l}{\n\c}
-C\ten{\c}{\m\n}C\ten{\l}{\l\c}  \eeqno{mamash} have particular
appeal; this tensor is symmetric under the interchange of the two
metrics. The scalars I use are then
 \beq \Up=\Gmn\Up_{\m\n},~~~~~\hat\Up= \hGmn\Up_{\m\n}.
 \eeqno{papash}
These are obtained from each other under metric interchange, and are
used symmetrically in constructing MTMS actions for
BIMOND.\footnote{The special appeal in the choice of the tensor
$\Up_{\m\n}$, and the two scalars constructed from it, is twofold
\cite{milgrom09}: First, it leads to a theory with a particularly
simple NR limit. Second, this choice is the best equivalent to the
choice of the Ricci tensor, $R_{\m\n}$, and the Ricci scalar,
$R=\Gmn R_{\m\n}$, in the Einstein-Hilbert Lagrangian of GR. In GR
we have to use the full curvature tensor, which contains the second
derivatives of the metric; this is both necessary (to get a scalar),
and harmless (because the second derivatives in the Einstein-Hilbert
Lagrangian do not affect the field equations). In fact, it is well
known that we can replace $R$ in the GR Lagrangian by $\Gmn
(\C\ten{\c}{\m\l}\C\ten{\l}{\n\c}
-\C\ten{\c}{\m\n}\C\ten{\l}{\l\c})$, exactly analogous to our
$\Up,~\hat\Up$ scalars. In our case we do not add the second
derivatives; this would be both harmful (as it would lead to a
higher-derivative theory), and unnecessary (as the first-derivative
terms are tensor in themselves).} The action terms $I\_M$ and $\hat
I\_M$ are the matter actions for standard matter and for the
putative TM (with their degrees of freedom denoted by $\psi\_i$ and
$\hat\psi\_i$), whose existence is suggested by the double metric
nature of the theory: Just as standard matter couples only to the
standard metric $\gmn$, TM couples only to the second metric
$\hgmn$. There is no direct interaction between the two matter
sectors. The dimensionless parameters $\b$ and $\a$ permit us to use
gravitational couplings in the two sectors, $G'=G/\b$ and $\hat
G'=G/\a$, that differ from each other and from $G$ itself. In the
rest of the paper I shall concentrate on symmetric versions of the
theory; this means taking $\a=\b$, and that the two metrics appear
in a symmetric way in the interaction term. As explained in
\cite{milgrom09,milgrom10}, there are good reasons to take
$\b\approx 1$ (see below). However, for the sake of generality I
shall keep a general value of $\b$ in some of the discussion. Note
that without the interaction term the theory consists of two
uncoupled copies of GR.
\par
The resulting field equations are of the form
 \beq  \b G\_{\m\n}+S\_{\m\n}=-8\pi G \Tmn,
\eeqno{misha}
 \beq \b\hat G\_{\m\n}+\hat S\_{\m\n}=-8\pi G \hTmn, \eeqno{nuvec}
where
 \beq G\_{\m\n}=R\_{\m\n}-\oot\gmn R,~~~~
 \hat G\_{\m\n}=\hat R\_{\m\n}-\oot\hgmn \hat R
   \eeqno{cutan}
are the Einstein tensors, and $\Tmn$, $\hTmn$ are the
energy-momentum tensors (EMTs) for the two sectors.
\par
The equations of motion of matter in the two sectors, in the
presence of gravity, are the standard ones since they are still
controlled by the same matter actions as in standard physics.
\par
The tensors $S\_{\m\n}$ and $\hat S\_{\m\n}$ fall from the variation
of the interaction term with respect to $\gmn$ and $\hgmn$,
respectively.\footnote{They satisfy, together, one set of four
Bianchi-like differential identities, stemming from the invariance
of the interaction term to general coordinate transformations
(details in \cite{milgrom09}).} The metrics appear in the
interaction in three ways: (i) through their derivatives, in
$C\ten{\a}{\b\c}$, (ii) contracting the $C\ten{\a}{\b\c}$ in the
quadratic scalars on which $\M$ depends, (iii) otherwise, such as in
the volume density $(g\hat g)^{1/4}$, or in scalars such as $g/\hat
g$ or $\Gmn\hgmn$. As a result, $S\_{\m\n}$ and $\hat S\_{\m\n}$
have the schematic form
 $$S_{\m\n}=(Q\{C\}\ten{\l}{\m\n})\cd{\l}
  +N\{CC\}_{\m\n}+
  \az^2 P\gmn,$$
  \beq \hat S_{\m\n}=(\hat Q\{C\}\ten{\l}{\m\n})\mcd{\l}
  +\hat N\{CC\}_{\m\n}+
  \az^2 \hat P\hgmn, \eeqno{gumshata}
with the three term types corresponding, respectively, to the above
types of appearance of the metrics. Here, $\{C\}\ten{\l}{\m\n}$
denotes tensors linear in  $C\ten{\a}{\b\c}$,\footnote{Examples of
such tensors are $C\ten{\l}{\m\n}$, $\d\ten{\l}{\m}C\ten{\a}{\a\n}$,
$\hgmn\Gab C\ten{\l}{\a\b}$, etc.} and $\{CC\}_{\m\n}$ denotes
tensors quadratic in them;\footnote{Examples of such tensors are
$C\ten{\c}{\m\l}C\ten{\l}{\n\c}$, $C\ten{\c}{\m\n}C\ten{\l}{\l\c}$,
$\Up\_{\m\n}$, etc.} they are the same in the two equations. Also,
``:'' denotes covariant derivation with respect to $\hgmn$.
$Q,~N,~P$, etc. are dimensionless, and presumably of order unity.
They depend on the two metrics through the quadratic scalars, and
possibly through scalars not containing the metric derivatives, such
as $g/\hat g$ or $\Gmn\hgmn$. $Q$ and $N$ are linear in derivatives
of $\M$ with respect to the quadratic scalars, e.g., $\Up$ and
$\hat\Up$ defined in Eq.(\ref{papash}), while $P$ gets contributions
from $\M$ itself and from its derivatives with respect to scalars
containing the metrics alone. This classification of the terms helps
understand some interesting special cases and limits of the theory.
In what follows I refer to these contributions as $Q,~N,~P$ terms,
respectively.
\par
From the symmetry of the problem we have
 \beq
\hat Q(\gmn,\hgmn)=-Q(\hgmn,\gmn), ~~~\hat
N(\gmn,\hgmn)=N(\hgmn,\gmn),~~~ \hat
P(\gmn,\hgmn)=P(\hgmn,\gmn).\eeqno{mmaa}
\par
BIMOND versions with $\b=1$ have a simple GR limit: Assume that in
the limit $\az\rar 0$, i.e., when the quadratic scalar arguments of
$\M$ go to infinity, $\M$ becomes independent of them. Then, the $Q$
and $N$ terms vanish. The $P$ terms then give CC-type terms that
might still couple the two metrics, but which are rather small
locally. It is also possible to choose $\M$ such that the coupling
disappears altogether in this limit, so the $P$ terms in $S\_{\m\n}$
collect to $\az\^2P(\infty)\gmn$ \cite{milgrom09}. The theory then
decouples into two copies of GR, possibly with a cosmological
constant, if $\M(\infty)\not = 0$. I have not yet been able to
ascertain if there is an acceptable GR limit for theories with
$\b\not =1$.\footnote{The NR limit of the theory does have a good
Newtonian limit even for $\b\not =1$.} This may be one reason to
prefer theories with $\b=1$.
\par
When the two metrics are equal, we have from Eq.(\ref{mmaa}) $\hat
Q=-Q\equiv -Q\_0,~\hat N=N\equiv N\_0,~ \hat P=P\equiv P\_0$, where,
furthermore, all these functions of the two metrics become
constants. This is because we cannot construct (nonconstant) scalars
from a single metric and its first derivatives, so all the scalar
arguments of these functions must become constants when
$\hgmn=\gmn$. Clearly, $Q$ terms vanish linearly in the metric
difference, and terms of type $N$ vanish quadratically. We are then
left only with $P$ terms, which are of a cosmological constant type
with $\Lambda\propto \az\^2 P\_0$. From this follows that, in
systems with identical matter and TM distributions, and the same
initial and boundary conditions, the metrics in the two sectors are
the same, and are a solution of the standard Einstein equations with
a cosmological constant (CC) $\Lambda$.\footnote{For example, a
double Schwarzschild solution is a vacuum solution of BIMOND, which
corresponds to a black hole of equal matter and TM masses.
Similarly, standard GR gravitational waves are also vacuum solutions
of BIMOND.}$^,$\footnote{Pictorially, in a double membrane picture
the Einstein-Hilbert actions for the two metrics may be viewed, as
usual, as the elastic energies of the membranes--the cost of
distorting them out of the flat. Then, the CC may be viewed as the
contact energy density, remaining when $\hgmn=\gmn$, and the rest of
the interaction is the ``cost of separating'' the membranes.}
\subsection{Cosmological background}
Preliminary considerations of cosmology in BIMOND have been
described in \cite{milgrom10} and in \cite{cz10}. As mentioned in
the Introduction, here I assume all along the special, but
attractive, case of an MTMS cosmology. This means that matter
sources in cosmology are assumed to be identical in the two sectors,
apart from differences in the initial seed inhomogeneities. If we
neglect these differences to lowest order, the MTMS cosmology in
BIMOND is identical to the Friedmann-Robertson-Walker (FRW)
cosmology in GR, with a CC, $\Lambda\propto\az\^2P\_0$, with
$G'=G/\b$ instead of $G$, and with the standard matter content. All
the standard results derived for the latter (Big Bang,
Nucleosynthesis, etc.) apply here as well (with $G'$ as
gravitational constant).
\par
To a higher order, the differences in the inhomogeneities in the two
sectors engender departure of the background cosmology from FRW: the
cosmological metric is no longer a solution of the Einstein
equations. This is discussed briefly in section \ref{back}. To some
approximation the cosmology could still be considered an FRW one,
but with nonstandard, ``phantom'' contributions to the cosmological
EMT in BIMOND. This sort of feedback goes beyond the feedback due to
departure from smoothness, which has been discussed extensively
respecting GR cosmology. Here it is specifically due to differences
between the inhomogeneities in the two sectors. As long as the
statistical properties of the inhomogeneities in the two sectors are
the same, the cosmological background metrics are still the same,
but they satisfy the BIMOND equations, instead of the Einstein
equations.
\par
The observed $^4{\rm He}$ abundance was used in \cite{carroll04} to
derive a bound of $|G'/G-1|\lesssim 0.13$, where $G'$ is the value
that applies at the formation time. This would imply, in the present
context, $|\b-1|\lesssim 0.13$, lending further motivation for
taking $\b=1$.
\subsection{\label{NR}The nonrelativistic limit}
Consider first the NR limit of BIMOND as applied to a system that is
small on cosmological scales (discussed in detail in
\cite{milgrom09}). Since we assume a symmetric cosmology, with the
two metrics equal, there is a coordinate frame in which locally the
two metrics are nearly Minkowskian. Because of local inequality of
matter and TM mass distributions, the departure from Minkowski is
not the same for the two metrics, which can thus be written as
 \beq \gmn=\emn-2\f\d\_{\m\n}+h_{\m\n},~~~~
 \hat g\_{\m\n}=\emn-2\fh\d\_{\m\n}+\hat h_{\m\n},
 \eeqno{rukun} with the potentials $\f,~\fh,~h_{\m\n},
~\hat h_{\m\n}$ treated to lowest order in the weak-field
approximation.\footnote{The separation into $\f$ and $h_{\m\n}$ is a
mere convenience: One defines $\f\equiv -(1+g\_{00})/2$, and
$h\_{00}=0$.}
\par
Note, importantly, that the weak-field approximation assumes that
$\f,\fh\ll 1$ (in our units where $c=1$). This, however, does not
imply that whenever $\f$ or $\fh$ appear they raise the order in the
perturbation: Unlike the case of GR, here we have an additional
dimensioned constant, $\az$, and quantities such as $\gf/\az$ or
$(\gf-\gfh)/\az$ are not assumed small, and do not raise the order
of terms in which they appear. In this light, $C\ten{\a}{\b\c}$ as
they appear in $\{C\}\ten{\l}{\m\n}$ and $\{CC\}_{\m\n}$ {\it are}
first order in the small parameters; however, in their appearance in
the arguments of $Q,~N,~P$, etc. $\az\^{-1}C\ten{\a}{\b\c}$ should
not be taken as small.\footnote{$\az$ may be viewed as a proxy for a
scale length $\ell\equiv c\^2/\az$. Terms such as $\gf/\az$ appear,
in fact, as $\ell\grad(\f/c\^2)$. The NR limit in the MOND context
has to be understood as $c\rar\infty$, $\ell\rar \infty$, with
$c\^2/\ell$ fixed.}
\par
The $N$ terms are second order and can be neglected. The $P$ terms
contribute to lowest order as a homogeneous CC, whose value we know
is observationally negligible in local NR physics. As to the $Q$
terms, since the $\{C\}\ten{\l}{\m\n}$ are already first order we
can equate the two metrics everywhere else with the background
Minkowski metric (so, for example, covariant derivatives become
normal derivatives). But note that this approximation applies only
to the metrics themselves, not to their derivatives, which appear
only in $\az\^{-1}C\ten{\a}{\b\c}$, which are not assumed small.

It can be seen\footnote{Following from Eq.(\ref{mmaa}), noting that
in this approximation $Q$ is symmetric to the interchange of the
metrics, as its argument is quadratic in their difference.} that to
the lowest order $Q=-\hat Q$, and so $S\_{\m\n}=-\hat S\_{\m\n}$.
Thus, the sum and difference of the two field equations are, in this
approximation,
 \beq \b(G\_{\m\n}+\hat G\_{\m\n})=-8\pi G (\Tmn+\hat\Tmn),~~~
 \b(G\_{\m\n}-\hat G\_{\m\n})+2 S\_{\m\n}=-8\pi G (\Tmn-\hat\Tmn).
 \eeqno{mipola}
The sum of the Einstein tensors is linear in the sums of the
potentials, and their difference, as well as $S\_{\m\n}$, depends
only on the differences of the potentials.
\par
In \cite{milgrom09} I treated the NR limit where one further assumes
that motions are slow, so that time derivatives are negligible
compared with spatial ones. I then showed that with the choice of
the quadratic scalar arguments given in Eq.(\ref{papash}), which I
shall assume,\footnote{The general form of the scalar was also
discussed in \cite{milgrom09}.} the field equations imply, in a
certain choice of gauge: $h_{\m\n}=\hat h_{\m\n}=0$. The two metrics
thus have  the form of the first-order metric in GR, but the NR
potentials, $\f$ and $\fh$, are determined not from the Poisson
equation but from field equations derived from the action
   \beq \L=-\frac{1}{8\pi
G}\{\b(\gfh)^2+\b(\gf)^2-\az^2\M[(\gf-\gfh)^2/\az^2]\}
  +\r(\oot\vv^2-\f)+
  \rh(\oot\hat\vv^2-\fh).  \eeqno{futcol}
This follows from the fact that to lowest order we have $\Up=\hat\Up
\propto (\gf-\gfh)^2$, and so the interaction term takes the form
$\az^2\M[(\gf-\gfh)^2/\az^2]$. (I use here the same symbol, $\M$,
for a function of a single variable: the reduction of the
relativistic interaction function, also denoted $\M$, which depends
on more variables.) This action also governs the particle equations
of motions in the two sectors, which take the standard form:
$\va=-\gf$ and $\hat\va=-\gfh$. Because the metric has the GR form
in terms of a single potential: $\gmn=\emn-2\f\d\_{\m\n}$, light
bending by slowly moving masses is expressed in terms of the
potential in the same way as in GR: photons ``see'' the same
potential as massive particles.
\par
The field equations for the potentials are
$$\D\f=\fpg\b^{-1}\r+\b^{-1}\div[(\gf-\gfh)\M']\equiv \fpg(\r+\rp),$$
\beq \D\fh=\fpg\b^{-1}\rh-\b^{-1}\div[(\gf-\gfh)\M']\equiv
\fpg(\rh+\rph),\eeqno{hugtal} where $\rp$ and $\rph$ play the role
of ``phantom matter'' (PM) densities for the two sectors.
\par
As detailed in \cite{milgrom10}, it is convenient to work with the
two potentials $\ft$ and $\fb$ such that
  \beq\f=\z\ft+\fb,~~~~~~~\fh=\z\ft-\fb, \eeqno{rabutret}
which satisfy
 \beq \D\ft=\fpg(\r+\rh),~~~~~
 \div\{\mt(|\gfb|/\az)\gfb\}=\fpg(\r-\rh),  \eeqno{katpet}
with the boundary conditions at infinity $\ft\rar 0$, while
$\fb\propto ln (r)$ when $M\not=\hat M$, and $\fb\rar 0$ when
$M=\hat M$; $M$ and $\hat M$ are the total masses in matter and TM
respectively. Here, $\z=(2\b)^{-1}$, and $\mt(x)\equiv
2\b-4\M'(4x\^2)$.\footnote{The factor $4$ in the argument accounts
for the fact that we use the variable $\fb=(\f-\fh)/2$.} MOND and
Newtonian behaviors for isolated, pure-matter systems, such as
galaxies, dictate the asymptotic behaviors $\mt(x\ll 1)\approx x$,
$\mt(x\gg 1)\approx (1-\z)\^{-1}$. Thus $\ft=\fN+\fNh$ is simply the
sum of the matter and TM Newtonian potentials, and $\fb$ is a MOND
potential satisfying the nonlinear Poisson equation proposed by
\cite{bm84} with $\r-\rh$ as a source (and a modified value of $\mt$
at large arguments).

\section{\label{problem}Fluctuations in an isolated MTMS system}
Consider first the development of weak fluctuations in a
cosmologically small, isolated, NR system. The two matter components
are treated as fluids. The time dependent problem is governed by the
NR continuity equations in the two sectors
  \beq \r\_t+\div(\r\vv)=0,~~~~~\rh\_t+\div(\rh\hat
  \vv)=0, \eeqno{cont}
and the respective Euler equations
 \beq
 \vv\_t+(\vv\cdot\nabla)\vv=-\gf-\r\^{-1}\grad p,
 ~~~~~\hat\vv\_t+(\hat\vv\cdot\nabla)\hat\vv=
 -\gfh-\rh\^{-1}\grad \hat p, \eeqno{euler}
where $p,~\hat p$ are the pressures in the two sectors.
\par
In a system made predominantly of matter--as we have, purportedly,
in  the solar system, galaxies, etc.--these equations lead to
standard MOND dynamics as they have been applied extensively in the
past. In this case, the MOND acceleration is generically larger than
the Newtonian one, and when the acceleration is smaller than $\az$,
dynamics is controlled by the MOND limit.
\par
However, as discussed in \cite{milgrom10}, things are rather
different in systems where the distribution of matter and TM are
close to each other (i.e., $|\r-\rh|\ll\r$). For example, when
$\r=\rh$ the dynamics is always Newtonian, no matter how small the
acceleration is.\footnote{This is the NR expression of the
reduction of BIMOND to GR for completely symmetric systems.}
\par
Consider then weak perturbations on an arbitrary, symmetric
background flow with $\r=\rh=\rz(\vr,t)$, $\vv=\hat\vv=\vvz(\vr,t)$.
Then, from Eq.(\ref{katpet}), $\fb=0$, and $\f=\fh=\b\^{-1}\fN\_b$,
where $\fN\_b$ is the Newtonian potential of $\rz$: we have a double
Newtonian system with an effective $G_e=G/\b$.\footnote{This case is
not to be confused with the Newtonian limit in a system made of pure
matter or pure TM, in which case $G$ itself plays the role of the
Newton constant.} If we seed the system with fluctuations that are
the same in the two sectors, they will remain so, with $\fb=0$ at
all times, and the development remaining quasi-Newtonian in both
sectors.
\par
Taking the lowest order in the fluctuations we have
 \beq \d\r\_t+\div(\d\r\vvz)+\div(\rz\d\vv)=0,
 ~~~~~\d\rh\_t+\div(\d\rh
  \vvz)+\div(\rz\d\hat\vv)=0, \eeqno{contper}
and the Euler equations
 $$\d\vv\_t+(\d\vv\cdot\nabla)\vvz+(\vvz\cdot\nabla)\d\vv=
 -\z\grad\d\ft
 -\gfb-\rz\^{-1}\grad \d p+\d\r\rz\^{-2}\grad \pz,$$
 \beq\d\hat\vv\_t+(\d\hat\vv\cdot\nabla)\vvz
 +(\vvz\cdot\nabla)\d\hat\vv=
 -\z\grad\d\ft+\gfb
 -\rz\^{-1}\grad\d\hat p+\d\rh\rz\^{-2}\grad \pz,
  \eeqno{eulerpre}
where $\pz$ is the pressure field in the unperturbed system (assumed
to be the same in the two sectors, as we assume the same equation of
state for the two). The potential fluctuations are obtained from:
 $$\D\d\ft=\fpg(\d\r+\d\rh), $$
 \beq\div(\mt|\gfb/\az|\gfb)=\fpg(\d\r-\d\rh).  \eeqno{gutna}
If $|\d\r|\sim|\d\rh|\sim|\d\r-\d\rh|$, the sources for these
equations are of similar magnitudes; but this need not
be so.
\par
It is useful to work with the sums and differences:
 \beq \vu=\d\vv+\d\hat\vv,~~~\bar\vu=\d\vv-\d\hat\vv,~~~
 q=\d\r+\d\rh,~~~\bar q=\d\r-\d\rh.\eeqno{mushta}
If we can assume that in both sectors $p$ can be taken a function of
$\r$; e.g., for a polytropic gas, or when the process is isentropic,
we write $\d p=(dp/d\r)(\rz)\d\r\equiv v\_s\^2(\rz)\d\r$ and get:
 \beq q\_t+\div(q\vvz)+\div(\rz\vu)=0, \eeqno{nuta}
  \beq \vu\_t+(\vu\cdot\nabla)\vvz+(\vvz\cdot\nabla)\vu=
 -2\z\grad\d\ft-\rz\^{-1}\grad(v\_s\^2  q)+q\rz\^{-2}v\_s\^2\grad
 \rz,
 \eeqno{nurala}
 \beq \D\d\ft=\fpg q, \eeqno{kitreq}
 \beq \bar q\_t+\div(\bar q\vvz)
 +\div(\rz\bar\vu)=0, \eeqno{contzol}
 \beq\bar\vu\_t+(\bar\vu\cdot\nabla)\vvz
 +(\vvz\cdot\nabla)\bar\vu=-2\gfb-\rz\^{-1}\grad(v\_s\^2
 \bar q)+\bar q\rz\^{-2}v\_s\^2\grad \rz, \eeqno{eulercot}
 \beq \div(\mt|\gfb/\az|\gfb)=\fpg\bar q.
 \eeqno{gutzol}
The barred and unbarred systems decouple, with the unbarred
fluctuations developing according to standard Newtonian theory,
while the barred fluctuations develope according to MOND. If we
start with $q$ and $\bar q$ fluctuations of a similar magnitude, and
the barred fluctuations are in the MOND regime, $\bar q$ grows
faster than $q$. This leads in the limit to $\d\r\approx -\d\rh$,
namely segregation of the two sectors. This echoes the fact that
matter and TM repel each other in the MOND regime \cite{milgrom10}.

\section{\label{cosmo}Cosmological fluctuation growth}
The use in the cosmological context of the above NR equations,
derived for an isolated system on a double Minkowski
background, can be justified for weak perturbations in NR matter, of
scale length much smaller than cosmological scales: Expand the
BIMOND field equations around their cosmological solution of the FRW
equations with a CC and $\hgmn=\gmn$. The EMTs are expanded around
their value for the homogeneous background with only the $00$
components contributing. For scales much smaller than the
(space-time) curvature radius the background metrics can be taken as
nearly Minkowski. Then, our formalism above applies.
\par
In this application, $\rz(t)$ is the ambient, homogeneous
cosmological density of the specific matter component we concentrate
on, and $\pz$ is its effective pressure.
\par
I shall take in what follows $\b=1$.
\par
As is standard, we express the space derivatives with respect to
comoving coordinates $\vR$ instead of the proper-distance
coordinates $\vr=a(t)\vR$, which appear in the above equations. We
also put $\vvz=H(t)\vr$, $\div{\vvz}=3H$, where $H=\dot a/a$ is the
Hubble parameter. Also, write the equations in terms of partial time
derivatives at a fixed comoving position: for the field $S$, $\dot
S\equiv S\_t+(\vvz\cdot\grad) S$. The continuity equation for the
background NR matter amounts to $\rz\propto a\^{-3}$, so,
$\dot\rz=\partial\rz/\partial t=-3H\rz$.
\par
Define $\e=\d\r/\rz$, $\eps=\eta+\hat\eta=q/\rz$,
$\bar\eps=\eta-\hat\eta=\hat q/\rz$. For these we have:
 \beq \dot\eps+a\^{-1}\nabla\_{\vR}\cdot\vu=0, \eeqno{nutaram}
  \beq \dot\vu+(\dot a/a)\vu=
 - a\^{-1}\grad\_{\vR}\d\ft-v\_s\^2 a\^{-1}\grad\_{\vR}\eps,
  \eeqno{hita}
 \beq \D\_{\vR}\d\ft=\fpg a\^2\rz\eps, \eeqno{kitram}
 \beq \dot{\bar \eps} +a\^{-1}\nabla\_{\vR}\cdot\bar\vu=0,
 \eeqno{contram}
\beq \dot{\bar\vu}+(\dot a/a)\bar\vu=
 -2a\^{-1}\grad\_{\vR}\fb-v\_s\^2 a\^{-1} \grad\_{\vR}\bar\eps,
 \eeqno{eulersup}
 \beq \nabla\_{\vR}(\mt|\grad\_{\vR}\fb/a\az|\grad\_{\vR}\fb)
 =\fpg a\^2\rz\bar \eps.
 \eeqno{gutzoba}
\par
The unbarred system is identical to that in standard gravity: the
perturbation sum develops fully in a Newtonian manner. Combining the
time derivative of the continuity equation with the $\vR$ divergence
of the Euler equation we get as usual
 \beq \ddot\eps+2(\dot a/a)\dot \eps=\fpg\rz\eps
+ v\_s\^2 a\^{-2}\D\_{\vR}\eps, \eeqno{kupopp} and for the space
Fourier components defined by $\eps=\int d\^3k\eps\_{\vk}(t)
e\^{i\vk\cdot \vr}$, we have the standard result
  \beq \ddot\eps\_{\vk}+2(\dot a/a)\dot \eps\_{\vk}=
  (\fpg\rz-v\_s\^2k\^2/a\^2)\eps\_{\vk}.
  \eeqno{kuptapi}
\par
The barred system, which is decoupled from the unbarred one,
describes the evolution of the perturbation difference; it behaves
as a single fluid governed by MOND, with the strength of the
governing potential, $\fb$, doubled [as expressed by the factor 2 in
front of $\gfb$ in Eq.(\ref{eulersup})], and with a Newtonian limit
for which the effective Newton constant is $G$.\footnote{Since
$\mt(\infty)=2$, $\fb$ is half the Newtonian potential sourced by
$\rz\bar\eps$ in this limit; the $1/2$ factor canceled by above
mentioned factor 2.} We can still eliminate the velocity $\bar\vu$
to get
 \beq \ddot{\bar\eps}+2(\dot a/a)\dot {\bar\eps}-
 2a\^{-2}\D\_{\vR}\fb -v\_s\^2
a\^{-2}\D\_{\vR}\bar\eps=0, \eeqno{kupluta} but here $\D\_{\vR}\fb$
is not simply expressible in terms of $\bar\eps$. The problem is
nonlinear even for the smallest of perturbations, and perturbations
on different scales can affect each other's growth (see below).
\par
Given some initial $\bar\eps(\vR,t\_0)$ and $\bar\vu(\vR,t\_0)$, we
can calculate $\dot{\bar\eps}(\vR,t\_0)$ from the continuity
equation, we then calculate $\fb$ from the potential equation, then
calculate $\ddot{\bar\eps}$ from Eq.(\ref{kupluta}), and so
propagate the system in time. The velocities are propagated in time
in parallel, using the Euler equation.

\subsection{\label{deep}The deep-MOND regime}
With matter and TM present, there are two senses in which we can
understand the term ``a deep-MOND system'': The weaker sense applies
to systems where $|\gfb|\ll \az$ so that $\fb$ is determined by
deep-MOND physics (with its peculiar symmetries, etc.). If we have
an inhomogeneity of proper length scale $\l$ that can be considered
in isolation, i.e., is not subject to an external-field effect (EFE)
from a larger-scale inhomogeneity of which it is a part (see below),
this amounts to
 \beq |\d\r-\d\rh| G\l\ll\az.  \eeqno{huplar}
In the stronger sense of the term, one requires, also, that
$|\gft|\ll|\gfb|$ so that the full potentials $\f$ and $\fh$ are
determined by deep-MOND physics. This requires further
 \beq |\d\r+\d\rh| G\l\ll(\az|\d\r-\d\rh|G\l)^{1/2}. \eeqno{miplaz}
In past applications of MOND, where TM is absent, the second
requirement follows automatically from the first. This is also the
case in the present context if $|\d\r-\d\rh|\sim
max(|\d\r|,|\d\rh|)$ (for example, if the fluctuations in the two
sectors are uncorrelated), but not if $|\d\r-\d\rh|\ll
max(|\d\r|,|\d\rh|)$.
\par
Because here we have full decoupling of the sum and difference
systems, with the latter being governed only by $\fb$, I shall use
the term in the weaker sense, which is enough to ensure that the
barred system is governed by deep-MOND physics.
\par
In the deep-MOND limit, Eq.(\ref{gutzoba}) is well approximated by
  \beq \nabla\_{\vR}(|\grad\_{\vR}\fb|\grad\_{\vR}\fb)=\fpg \r\_0
\az\bar \eps, \eeqno{gutmona} where $\r\_0=a\^3\rz$. This equation
is conformally invariant \cite{milgrom97,milgrom10}.
\subsection{Coupling of inhomogeneities on different scales}
The implications of Eq.(\ref{gutzoba}) for the dynamics in massive
systems have been discussed extensively (e.g., in
\cite{bm84,milgrom02}). One clear outcome of these studies concerns
the way in which the dynamics of a larger, mother system couple to
those in a smaller subsystem. When the two do not differ much in
size and/or mass, the interaction between them, which results from
the nonlinearity of the theory, is not easy to describe. However,
when the subsystem is much smaller, and much less massive, than the
mother system (such as a for star in a galaxy, or a galaxy in a
galaxy cluster), the interaction is simply described: The subsystem
moves within the mother system as if it where a test particle,
irrespective of its own intrinsic accelerations. In other words, the
dynamics of the mother system is oblivious to the internal dynamics
in its small subsystems, which can be treated as test particles
moving in their combined mean field. On the other hand, the dynamics
within the subsystem may be greatly affected by the acceleration,
$g\_{ex}$, with which it falls within the mother system, provided
this acceleration dominates the internal ones, and is in the MOND
regime ($g\_{ex}\lesssim \az$). The intrinsic dynamics is then
quasi-Newtonian with an effective gravitational constant $G'\sim
G\az/g\_{ex}$, and some mild anisotropy induced by the external
field. This is known as the MOND external-field-effect (EFE).
\par
In the present context, the mass source in the equation for the
potential, $\rz\bar\eps$, must exhibit, at any time, a continuum
spectrum of fluctuations, and it is difficult to describe the mutual
couplings of the different scales. We expect, however, that the
evolution of a fluctuation in $\bar\eps$ on a given scale, will be
affected mainly by larger-scale perturbations within which it is
nested, in the spirit of the EFE.
\par
A difference fluctuation of magnitude $\bar\eps$ and (proper) scale
length $\l$, would have a characteristic intrinsic acceleration,
when in isolation, $\sim (\az G\rz\bar\eps \l)\^{1/2}$. Thus, if,
e.g., $\bar\eps\propto \l\^{-\a}$, for $\a>1$ the typical
acceleration decreases with increasing scale, and we do not expect a
pronounced EFE. For $\a<1$, the largest scales determine the
dynamics within smaller ones, as described roughly by the EFE. For
$\a=1$ the accelerations on all scales are of the same order.

\subsection{Pressure vs gravity}
For the unbarred, fluctuation-sum system, the standard (Jeans)
criterion tells us when pressure becomes unimportant in impeding
gravitational growth. This applies [e.g., per Eq.(\ref{kuptapi})],
when $\rz G\l\^2\gg v\_s\^2$, or, equivalently, when $M\_b\gg
G\^{-3/2}v\_s\^3\rz\^{-1/2}$, where $M\_b=\rz\l\^3$ is the
background mass (in either sector) in a cube of side $\l$.
\par
For the barred system, the Jeans criterion for gravity dominance is
very different. We saw that because of the nonlinearity of this
system we cannot consider separately the growth of a single
perturbation with a given amplitude and scale length, as all modes
are, in principle, coupled. I discuss two end cases. First, consider
a perturbation of magnitude $\bar\eps$ and (proper) scale $\l$ that
is not affected by larger-scale overdensities (through an analog of
the EFE). From Eq.(\ref{eulersup}), or Eq.(\ref{kupluta}), the Jeans
criterion is still $|\gfb|\gg v\_s\^2|\grad\bar\eps|$. From
Eq.(\ref{gutzoba}) we now have in the MOND limit
$|\gfb|\^2/\l\az\sim G\rz|\bar\eps|$. Put together, these give the
condition for gravity dominance
 \beq M\_b
\gg (G\az)\^{-1}v\_s\^4|\bar\eps| \eeqno{mipoga} (where gradients
are substituted by division by $\l$;
e.g.,$|\grad\bar\eps|\sim|\bar\eps|/\l$).
\par
Note, in particular, the appearance of $|\bar\eps|$ in the
criterion, in addition to the expected characteristic, deep-MOND
$M\propto v\^4$ relation. This occurs because the pressure increment
is linear in $\bar\eps$, while the potential increment $\fb$ is
proportional to $|\bar\eps|\^{1/2}$. This $|\bar\eps|$ factor makes
the criterion much easier to satisfy for weak
perturbations,\footnote{In \cite{milgrom89} I gave the Jeans
criterion for a MOND universe as $M\_b \gg (G\az )\^{-1}v\_s\^4$.
This was based on a pure matter world-picture, where it was assumed,
tentatively, that the source for the MOND potential includes the
background density, and so the potential increment is linear in the
density increment. Here, with an MTMS background, the background
MOND potential vanishes, so the perturbation alone sources $\fb$,
and so $|\gfb|\propto|\bar\eps|\^{1/2}$.}
\par
If the particular perturbation at hand cannot be considered in
isolation, but is subject to an EFE characterized by $g\_{ex}$, the
Jeans criterion takes the form $\rz G(\az/g\_{ex})\l\^2\gg v\_s\^2$,
or
 \beq M\_b\gg G\^{-3/2}(g\_{ex}/\az)\^{3/2}v\_s\^3\rz\^{-1/2}.
 \eeqno{mirteq}
\par
As a concrete example, consider a baryonic, difference perturbation,
of magnitude $\bar\eps$ and scale $\l$, that may be considered
unaffected by an EFE. Taking the ambient temperature appropriate for
the matter-dominance era $T\approx 3\times
10\^3(a/10\^{-3})\^{-1}{\rm K}$, and ambient density $\rz\approx
4\times 10\^{-22}(a/10\^{-3})\^{-3}\gcmt$, the condition
(\ref{huplar}) for the perturbation being in the deep-MOND regime
can be written
 \beq M_b/\msun\ll 2\times 10\^7\left(\frac{a}{10\^{-3}}\right)\^{6}
|\bar\eps|\^{-3}.   \eeqno{julapa} The condition (\ref{mipoga}) for
gravity dominance over pressure is
\beq M_b/\msun\gg 3\times
10\^5\left(\frac{a}{10\^{-3}}\right)\^{-4}|\bar\eps|.
     \eeqno{julagat} In
comparison, the Jeans criterion relevant for the sum of the
perturbations is
 \beq M\_b/\msun\gg 10\^6 \left(\frac{a}{10\^{-3}}\right)\^{-1.5}.
  \eeqno{cupom} So, there is a wide range of
scales for which pressure stymies the growth  of $\eps$ but not of
$\bar\eps$.

\subsection{\label{later}Later stages} The analysis above highlights
the way in which small-scale, NR density inhomogeneities grow when
they are still small compared with the ambient density. We have
learned much about the mechanisms that are at play and how they
greatly differ from those underlying perturbation growth in standard
dynamics. For example, we noted the shepherding of matter
overdensities by flanking TM overdensities (and vice-versa): a
matter overdensity grows faster when it sits in a TM depression. A
related effect is the segregation of matter and TM, as evinced by
the tendency of difference fluctuations to grow faster than the sum
\par
But, beyond these qualitative observations, a numerical attack on
the problem is needed to follow the growth of even weak
perturbations in detail, because of the inherent MOND nonlinearity.
This is also, clearly, the case when we deal with structure buildup
to its presently observed stage, where the perturbations are not
small. As mentioned in the introduction there have been several
$N$-body studies of structure formation in MOND, and these could be
easily adapted to include TM.
\par

\par
Without numerical studies, it is difficult to tell exactly how, and
to what extent, these trends continue to shape structure in the
later stages. Some questions that naturally arise are: ``Up to what
scales are matter and TM effectively segregated today?'' In
\cite{milgrom10}, I conjectured that this happens on the
supercluster-void scales; namely, that matter and TM have formed
interlacing cosmic webs in which TM concentrations sit in the
observed matter voids, and vice-versa, and matter and TM filaments
avoid each other. Another important question is: ``How rare are TM
objects, such as galaxies, in matter territory, and what observable
effects can they have?'' If the emptiness of matter voids is an
indication, we do not expect TM galaxies within large matter
concentrations. Other questions are: ``Does TM help empty matter
voids more efficiently than in the CDM paradigm?'' (For possible
indications for unexpectedly large and empty voids see
\cite{peebles01,tully07,tullyetal08,tik09,
tiketal09,weygaert09,peebnuss10}.) ``Are higher very-large-scale
velocities produced in the present picture than in the framework of
CDM?'' (See \cite{watkins09,lavaux10,kashlinsky10} for possible
indications of such unexpectedly high velocities.) ``Are objects
with high peculiar velocities more common in our world picture than
in the CDM paradigm?'' (See \cite{angus08} for a discussion of the
issue in connection with the bullet cluster, and \cite{lk10} for a
recent assessment, and for references to earlier work.)
\par
We can get some guidance in pondering over these questions by
considering the interaction between well separated matter and TM
bodies. This question has been considered in detail in
\cite{milgrom10}, which deals, in a sense, with the opposite of the
situation we treat here of small departures from well mixed
configurations.
\par
The basic result found there is that matter and TM bodies repel each
other in the MOND regime, and (for $\b=1$) they do not interact in
the high-acceleration regime. For example, using the conformal
invariance of the deep-MOND limit of the theory, I calculated the
force between two masses $M\_1$ and $M\_2$ in the deep-MOND regime
to be (defined as negative for attraction)
 \beq F=-\frac{2}{3}
 \frac{(\az G)\^{1/2}}{r}[(M\_1\pm M\_2)\^{3/2}
 -M\_1\^{3/2}-M\_2\^{3/2}], \eeqno{mitaq}
where the plus sign applies when the two masses are in the same
sector (in which case $F$ is always attractive), and the minus sign
when they are different (in which case the force is always
repulsive); $r$ is the separation. We see that for equal masses the
repulsive force between matter and TM masses M is $(4/3)(\az
G)\^{1/2}M\^{3/2}/ r$, which is $\approx 2.4$ times larger than the
attractive force between two masses $M$ in the same sector. This can
give us an idea of how far from each other separated bodies of
matter and TM can separate over the Hubble time.\footnote{This is
only a rough estimates, since such bodies are hardly ever isolated
on the relevant scales.}
\par
The characteristic separation speed for matter and TM masses $M$ is
$\sim 2.3(\az G M)\^{1/4} \sim 260(M/10\^{10}\msun)\^{1/4} \kms$. In
a Hubble time his produces a separation of several megaparsecs for
galactic mass objects and a few tens of megaparsecs for
galaxy-cluster mass objects. A test mass in one sector in the (MOND)
field of a mass $M$ of the other sector has typical separation
speeds about half of that given above: a massive TM cluster can
shoot out a matter bullet with a speed of several thousand
$\kms$.\footnote{If the bullet test mass starts at rest a distance
$r\_0$ from a TM mass M, its speed at a distance r is, assuming the
system is isolated, $v(r)=(4\az G M)\^{1/4}ln\^{1/2}(r/r\_0)\approx
1.6\times 10\^3(M/10\^{14}\msun)\^{1/4}ln\^{1/2}(r/r\_0) \kms $.}
\par
Another potentially useful result for a late-stage system made of
separate matter and TM bodies is the general, `virial' relation that
holds exactly in the deep-MOND regime: For a system made of matter
and TM pointlike masses $m\_i$ and $\hat m\_j$, respectively, at
positions $\vr\_k$ ($k$ running over all masses), we have
 \beq \sum\_k \vr\_k\cdot\vF\_k=-\frac{2}{3}
 (\az G)\^{1/2}[|\sum\_i m\_i-\sum\_j \hat m\_j|\^{3/2}
 -\sum\_i m\_i\^{3/2}-\sum\_j \hat m\_j\^{3/2}],   \eeqno{luprat}
where $\vF\_k$ is the force on particle $k$. (The choice of origin
is immaterial as $\sum\vF\_k=0$.) For a system with equal total
masses, the first term drops, which gives an expression that is
always positive; this means that on average the forces are pointing
outward, bespeaking a mean repulsion.\footnote{Such a system does
not satisfy a kinematic virial relation: since
$\vF\_k=m\_k\dot\vv\_k$, we find that $d(\sum\_k
m\_k\vr\_k\cdot\vv\_k)/dt>0$; so no steady state can be reached.}
For example, in a system  made of $N$ equal matter masses, $m$, and
$N$ equal TM masses, $m$, we get
 \beq \sum\_{k=1}\^{2N}\vr\_k\cdot\vF\_k=\frac{4}{3}(\az
G)\^{1/2}Nm\^{3/2}. \eeqno{viva} Thus, for the accelerations we have
 \beq \langle \vr\_k\cdot\va\_k\rangle=\frac{2}{3}(\az Gm)\^{1/2}.
   \eeqno{nutas} The acceleration thus points outward on
average, implying global self repulsion. The magnitude of the
typical acceleration is $\sim (\az Gm)\^{1/2}/R$, where $R$ is the
characteristic system size. This is smaller by a factor $\sim
N\^{-1/2}$ than the typical MOND acceleration in a pure-matter
system of the same mass. Note, however, that this mean acceleration
is only relevant if the matter and TM masses are very well mixed.
If, in fact, they form separate matter and TM clumps within the
system, as they would tend to do, these clumps would separate with
higher accelerations.

\section{\label{back}Phantom matter and its back
 reaction on cosmology}
There has been much discussion recently regarding back reaction of
inhomogeneities on cosmological evolution in GR. The issue concerns
the extent to which we err by applying FRW theory (which assumes
homogeneous energy density) to the mean cosmological energy density,
instead of calculating cosmological evolution correctly with
inhomogeneously distributed energy, and then averaging the outcome.
(For recent discussion of this issue, with references to earlier
work, see, e.g., \cite{clifton09,clifton10}.) The exact importance
of this effect is still moot.
\par
Obviously, such effects exist in BIMOND as well, but they are not
the subject here. In a MTMS universe governed by BIMOND there are
added back-reaction effects of inhomogeneities, which, as we saw,
induce the appearance of phantom matter components that should have
influence on cosmology.
\par
In a universe with different matter and TM inhomogeneities, the
$S\_{\m\n}$ and $\hat S\_{\m\n}$ terms in the BIMOND field equations
(\ref{misha})(\ref{nuvec}) do not vanish; so, what is their effect
on the cosmological background? Strictly speaking, we then have to
solve the full BIMOND cosmological equations. However, to some
approximation, we may treat the resulting departure from GR
perturbatively, by viewing the effects of these terms as
perturbations on the MTMS, FRW cosmology: In this approximation, the
$S\_{\m\n}$ and $\hat S\_{\m\n}$ terms are viewed as the EMTs of
MOND phantom matter that can effectively be added to $\Tmn$ and
$\hTmn$ in sourcing cosmic gravity. We saw that to zeroth order in
the metric difference, all that remains from these terms is a CC
term, which has to be reckoned with in zeroth-order cosmology. This
cosmology is then used to calculate the development of perturbations
in the two sectors, starting from some seed perturbations. The next
stage in such an approximation scheme would be to use the deduced
local departures of the two metrics from the FRW metric, used in the
lower order approximation, to calculate local contributions to
$S\_{\m\n}$ and $\hat S\_{\m\n}$, and put them back as sources in
the equations for cosmology. This program clearly deserves a
thorough treatment, which, however, is beyond our scope here. In
particular, the results of such an analysis is expected to depend on
the details of the specific BIMOND version at hand, and there is a
large variety of versions to explore. Here I only describe briefly
some general properties of the different term types defined in
Eq.(\ref{gumshata}), as they would be fed back into the cosmological
equations.
\par
The zeroth order of the $P$ term (in the matter sector),
$\az\^2P\_0\gmn$ gives a genuine CC term [if $P\_0\not = 0$]. When
matter and TM separate, $P$ becomes space-time dependent, and in
this regard behaves as a ``dark energy'' field. In the weak-field
approximation, we can substitute in the variables on which $P$
depends $\gmn\approx\hgmn$, but $\Up\approx\hat\Up\propto
(\gf-\gfh)\^2$. The contribution of the $P$ term is then
 \beq S\^P\_{\m\n}\approx\az
P(z)\gmn, ~~~~~~~z\equiv (\gf-\gfh)\^2/\az\^2.  \eeqno{mimipa} From
Eq.(\ref{mmaa}) we see that $\hat P=P$ in this approximation. The
value of $P$ in this expression varies in space-time between $P\_0$
and $P(\infty)$. (As explained in \cite{milgrom10}, considerations
of the GR limit of the theory, for  $\az\rar 0$, dictate that
$P(\infty)$ is not infinite.) Today we have in much of the volume of
the universe $z\lesssim 10\^{-3}$; so in much of the volume
$P\approx P\_0$, but in galaxies or in galaxy clusters (within about
a megaparsec of the center), $z\gtrsim 1$, and near stars $P\approx
P(\infty)$. Since the values of $P$ around matter concentrations are
the same as its values around similar TM concentrations
(interchanging matter and TM amounts to interchanging $\f$ and $\fh$
to which $P$ is invariant), the large-scale average of $P$ gives a
time dependent effective CC, $\Lambda (t)\propto\langle
P\rangle\_{space}\not =P\_0$.
\par
The next order contribution, which appears with the separation of
matter from TM, comes from the lowest order in the $Q$ terms, whose
contribution to the cosmological density ``phantom'' EMT is
 \beq \Tmn\^{(p)}=
 (8\pi G)\^{-1}(Q\{C\}\ten{\l}{\m\n})\cd{\l},
   \eeqno{rushpa}
 giving in the weak-field approximation
\beq \Tmn\^{(p)}\approx -\hTmn\^{(p)}\approx
 (8\pi G)\^{-1}(Q\{C\}\ten{\l}{\m\n})\_{,\l}.
   \eeqno{rushpara}

In the weak-field-slow-motions approximation, where time derivatives
are neglected, only the $00$ component contributes and gives,
simply, the totality of all the PM gotten from Eq.(\ref{hugtal}),
namely a NR PM density
 \beq \rp=(\fpg)\^{-1}\div[(\gf-\gfh)\M'(z)]. \eeqno{jupapa}
In a MTMS universe this density averages to zero over large
cosmological volumes. This can be seen in two (independent) ways.
First, it follows from the fact that under the interchange of matter
and TM, $\Tmn\leftrightarrow\hTmn$, we have
$\gmn\leftrightarrow\hgmn$. In the NR limit this means that
$\r\leftrightarrow\rh$ leads to $\f\leftrightarrow\fh$, which, in
turns gives $\rp\rar -\rp$: The phantom density around a matter
concentration is equal and opposite in sign to that around a similar
TM concentration. In our world picture the large-scale averages of
the universe are invariant to matter-TM interchange, hence the
average phantom density, which changes sign, must vanish. This
result holds also for a finite system with matter-TM symmetry. For
example, in a system where matter-TM interchange can be achieved by
a volume-preserving transformation (such as translation, reflection,
rotation) the total phantom mass in a symmetric volume vanishes.
\par
We can also argue differently, noting that the phantom density is a
divergence of a vector $(\gf-\gfh)\M'(z)$, whose absolute value is
bounded (outside of point masses). As a result, its volume average
vanishes in the limit of a large volume. This property of the
phantom density follows from the form of the relativistic $Q$ term
contribution: We see from Eq.(\ref{rushpa}) that the trace of the
phantom EMT is a covariant divergence of a vector $V\^{\l}\propto
Q\Gmn\{C\}\ten{\l}{\m\n}$. Its integral over a space-time volume is
hence a surface integral.
\par
Even if the large-scale average of the NR phantom density averages
to zero it could still affect cosmology due to its inhomogeneity. In
addition, there may be higher order contributions that do not
average to zero.
\par
The $Q$ terms may also have contributions of the form $(\M'\Gab
C\ten{\l}{\a\b})\cd{\l}\gmn$, or $(\M'\Gab
C\ten{\l}{\a\l})\cd{\b}\gmn$, which are of the form
$V\^{\l}\cd{\l}\gmn$; i.e., they contribute in cosmology as dark
energy with oscillating, vanishing-average density.
\par
The PM around matter concentrations can be (and is) observed, and is
interpreted by most as ``dark matter.'' As discussed in
\cite{milgrom10}, the PM around TM  concentrations may be observed
with weak lensing of matter photons (it does not produce strong
lensing), on which it acts as a repelling mass.
\par
For the $N$ terms, the contribution of lowest order in the departure
from metric equality is the second order. In the NR limit their
characteristic contribution to the phantom EMT is of the order of
$(8\pi G)\^{-1}zN(z)(\az/c)\^2$ [$z$ defined in Eq.(\ref{mimipa})].
Using the well known proximity $\az\approx cH\_0/2\pi$, we can write
this contribution as $\sim 10\^{-2}zN(z)\r\_c$, where $\r\_c$ is the
critical density today. We saw above that in most of the
cosmological volume today $z\lesssim 10\^{-3}$. Also we have that
$N(0)$ is of order unity (from considerations of the MOND limit, see
\cite{milgrom09,milgrom10}), and $N(\infty)=0$ (from considerations
of the GR limit of BIMOND). $N$ is related to the extrapolating
function of MOND and is constrained to remain of order unity, or
below, for all arguments. So the contribution of the $N$ terms to
the density can be at most of order $\r\_c$ (in systems where $
|\gfb|\sim\az$), but some orders of magnitude smaller than $\r\_c$
in most of the cosmic volume. It would be interesting to see what
their tensorial form is, and to what extent they produce observable
lensing effects.
\par
Note, finally, that inasmuch as the statistical properties of matter
in the two sectors are the same, so are the feedback effects
discussed above.\footnote{Even if the initial fluctuations are
different in the two sectors, we saw that their statistical
properties are driven to equalization by the faster, MOND-like
growth of the fluctuation difference.} Namely, the effects of
inhomogeneities, and the statistical properties of the PM (their
cosmological mean densities, power spectra, etc.) are the same in
the two sectors. The two mean background metrics thus remain the
same.

\section{\label{discussion}Discussion}
I have considered the dynamics of small, NR perturbations in an MTMS
universe. An intriguing aspect of this world picture is that MOND
dynamics, cosmological or local, are completely absent in a
perfectly symmetric universe, apart from a possible appearance of a
CC of order $(\az/c)\^2$. All other departure from standard physics
encapsulated in MOND--as it appears in cosmology, in structure
formation, and in bound systems, such as the solar system or
galactic systems--is entirely an effect of ``separation'' between
matter and TM, or separation between the two metrics. In a
double-membrane picture of BIMOND, MOND effects are due to local
separation of the membranes.
\par
Twin matter plays two roles in our narrative: In the first, it
provide a counteracting medium to matter; similar to the role of a
neutralizing positive ion background to an electron gas. As a result
of this presence, we could assume the very appealing symmetric
cosmology, identical to that of GR with a CC. In addition, the
presence of matter and TM in equal average quantities leads to a
well-defined MOND dynamics, in which the MOND potential is sourced
only by departures from uniformity, not by the bulk. To fulfill this
role, TM only has to be present in the same amount as matter; it
does not necessarily have to clump.
\par
In its second role, TM plays an equivalent part to that of matter;
so it can be likened to a positron component in a globally neutral
electron-positron plasma (with mutual repulsion, and no
annihilation). In this role it also takes part in shaping the
large-scale structure of matter.
\par
There are several qualitative features that make the growth of
fluctuations in the proposed world picture different from the
standard one. Among them: (i) Two initial fluctuation spectra, not
one, determine the large-scale structure at later times. For the NR
problem discussed here, we can think of these as the fluctuations in
matter alone, and of those in the matter-TM density difference. This
is important to appreciate, for example, because we have strong
constraints on matter fluctuations at the time of matter-radiation
decoupling, from the CMB anisotropies; but, we have none on the
difference fluctuations $\bar\eps$, at that time. So, numerical
simulations have to explore all possible options. (ii) The NR
fluctuations in the sum and difference of the densities form two
decoupled systems controlled by different types of gravitational
physics. The sum system is underlaid by standard Newtonian gravity,
while the difference system is underlaid by MOND gravity, which is
stronger for the same source strength. (iii) In the difference
system, the competition between gravity and pressure is strongly
tilted in favor of gravity, with the Jeans criterion being much more
lenient than in the sum system. (iv) As a result of (ii) and (iii),
the difference fluctuations tend to grow faster than the sum
fluctuation, leading towards a limit where $|\bar\eps|\gg|\eps|$, or
$\d\r\approx -\d\rh$. This implies tendency towards anticorrelation
of matter and TM overdensities, or to segregation of the two
sectors. This applies not only when the perturbations are weak: the
fact that distinct matter and TM bodies repel each other for low
accelerations implies that the process of segregation continues
during the stages where the overdensities are high. (v) There is an
added enhancement, due to the fact that once the fluctuations in the
two sectors are anticorrelated, with maxima of each residing in
minima of the other, each overdensity grows not only by the pull of
its own gravity, but also by shepherding by the repulsive action of
the flanking overdensities in the other sector. (vi) There can be
strong coupling between the perturbations on different scales due to
the nonlinearity inherent in MOND.
\par
There is much that remains to be studied, even within the narrow
paradigm of the MTMS universe. An even richer prospect opens if one
considers more general cases such as matter-TM asymmetric BIMOND
versions, or asymmetric universes in which the two sectors are not
equally populated, or do not have exactly the same properties.
\par
For example, we do not know why a universe that presumably started
with a vanishing baryon-number density has ended up having more
baryons than antibaryons, with the difference resulting in the
present-day baryon density. This baryogenesis, as it is called,
which requires a breakdown of symmetries (CP and baryon-number
conservation), is thought to have occurred during some early phase
of thermal disequilibrium. If the corresponding conditions in the
two sectors where slightly different, we could have ended up with
different baryon densities in the two sectors. It would be
interesting to study the effects of such departure from MTMS, to
lowest order in the density difference, to see what effects it has
on both the cosmological evolution, and on structure growth.

\section*{Acknowledgements}

This research was supported by a center of excellence grant from the
Israel Science Foundation.

\clearpage

\end{document}